\documentclass[a4paper,twocolumn,superscriptaddress,11pt,accepted=2018-12-08]{quantumarticle}
\pdfoutput=1
\usepackage[utf8]{inputenc}
\usepackage[english]{babel}
\usepackage[T1]{fontenc}
\usepackage{amsmath}
\usepackage{amsfonts}
\usepackage{amssymb}
\usepackage[numbers,sort&compress,nonamebreak]{natbib}
\usepackage{graphicx}
\usepackage{hyperref}

\newcommand{\mb}{\mathbf}
\newcommand{\bs}{\boldsymbol}
\newcommand{\T}{\text}
\newcommand{\rme}{\mathrm{e}}

\begin{document}

\title{Quantum Interference of Force}

\author{Raul Corr\^ea}\email{raulcs@fisica.ufmg.br}
\affiliation{Departamento de F\'isica, Universidade Federal de Minas Gerais, Caixa Postal 701, 30161-970, Belo Horizonte, MG, Brazil}
\orcid{0000-0001-6624-3759}
\author{Marina F. B. Cenni}
\affiliation{Departamento de F\'isica, Universidade Federal de Minas Gerais, Caixa Postal 701, 30161-970, Belo Horizonte, MG, Brazil}
\author{Pablo L. Saldanha}\email{saldanha@fisica.ufmg.br}
\affiliation{Departamento de F\'isica, Universidade Federal de Minas Gerais, Caixa Postal 701, 30161-970, Belo Horizonte, MG, Brazil}
\orcid{0000-0002-1844-0771}


\begin{abstract}
We show that a quantum particle subjected to a positive force in one path of a Mach-Zehnder interferometer and a null force in the other path may receive a negative average momentum transfer when it leaves the interferometer by a particular exit. In this scenario, an ensemble of particles may receive an average momentum in the opposite direction of the applied force due to quantum interference, a behavior with no classical analogue. We discuss some experimental schemes that could verify the effect with current technology, with electrons or neutrons in Mach-Zehnder interferometers in free space and with atoms from a Bose-Einstein condensate.
\end{abstract}



\maketitle


\par Interference phenomena on quantum systems can lead to extremely curious physics. The most canonical example is the double-slit experiment, which according to Feynman ``has in it the heart of quantum mechanics'' \cite{feynman3,tonomura89,bach13}. Other intriguing examples are delayed-choice experiments \cite{wheeler78,jacques07,manning15,vedovato17}, quantum erasers \cite{scully91,herzog95,durr98,walborn02}, ``interaction-free'' measurements \cite{elitzur93,kwiat95,peise15}, and quantum delayed-choice experiments \cite{ionicioiu11,peruzzo12,kaiser12}, among many others. 

Here we discuss another very curious quantum interference effect inspired by a recent work from Aharonov et al. \cite{aharonov132}. In their article, the authors show that the classical limit of quantum optics is achieved in a strange way when we look at how photons transfer momentum to a mirror inside an interferometer, if the mirror position is treated quantum-mechanically. They showed that, depending on at which port of the interferometer the photon exits, it may transfer to the mirror an average momentum in the opposite direction that one would expect, due to an interference effect. This can be seen as a quantum random walk in the momentum space \cite{aharonov93}, in which the interferometer beam splitters play the role of a quantum coin that decides whether the mirror will be pushed by the photon or not, while under suitable post-selection it ends up being pulled.

We generalize this result upon considering anomalous momentum transfers to general quantum objects and by considering that the momentum transfer in one of the arms of the interferometer may be of the same order of magnitude than the initial momentum uncertainty of the quantum object, not necessarily in the weak interaction regime considered in Ref. \cite{aharonov132}. These generalizations lead us to some proposals for feasible experiments that could observe an anomalous momentum transfer to a quantum object due to the quantum interference of force. Some of them use quantum particles (electrons or neutrons) propagating in a Mach-Zehnder interferometer in free space, with a force that acts in only one of the interferometer arms. Another one uses an interferometer based on the internal degrees of freedom of atoms from a Bose-Einstein condensate, with a force that acts differently on these internal states. With an appropriate post-selection, the quantum superposition of a positive force with no force in an ensemble of quantum particles may generate a negative average momentum transfer to these particles. No classical system can present this behavior, such that the quantum interference of force that we discuss here is a genuinely quantum effect.

\begin{figure}[t]
  \centering
    \includegraphics[width=8.5cm]{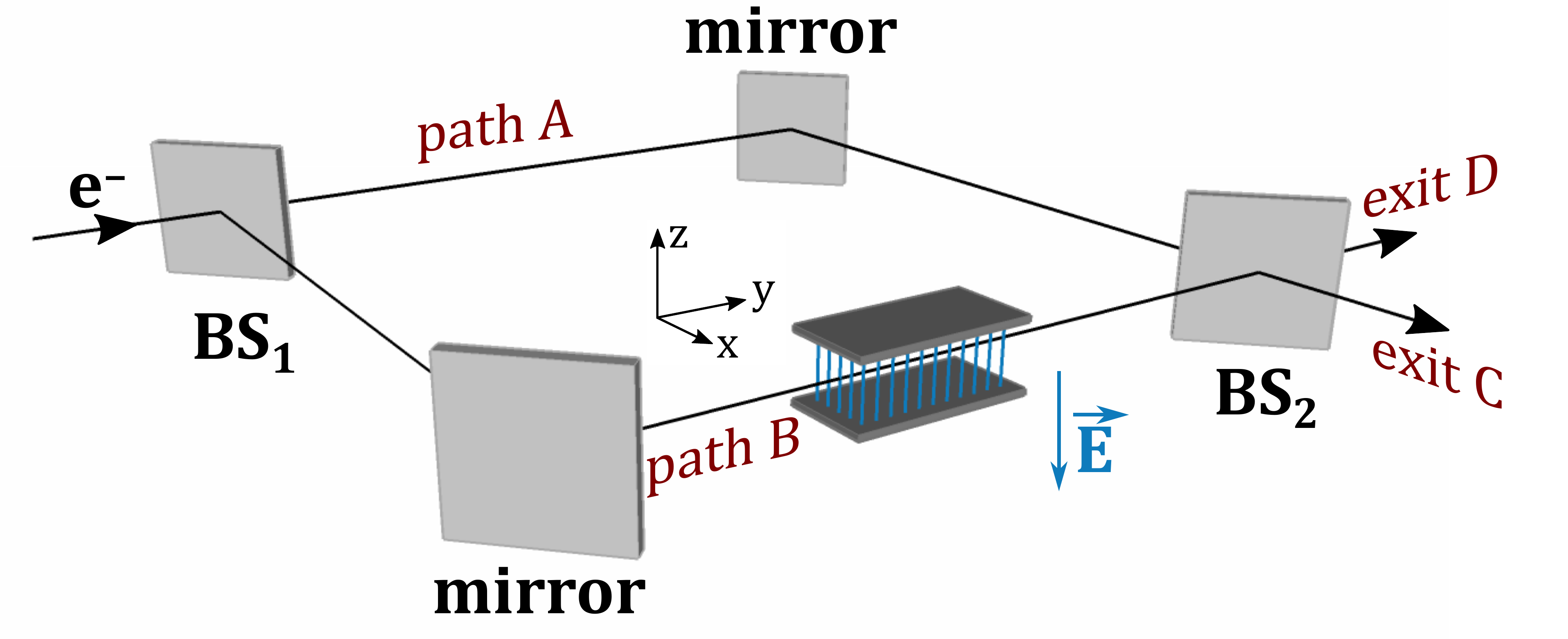}
  \caption{Simplified scheme of an electron Mach-Zehnder interferometer. The first beam splitter (BS$_1$) splits the electron wave function into two path components A and B. Path B contains a capacitor with an uniform electric field $\mb{E}$ in the $-\bs{\hat{z}}$ direction, generating a force in the $\bs{\hat{z}}$ direction on a propagating electron. Mirrors re-join the wave function components and mix them at a second beam splitter (BS$_2$), whose exit ports are labeled C and D.}\label{fig:interferometer}
\end{figure}

To discuss the effect, let us consider a quantum particle propagating through a two-paths Mach-Zehnder interferometer. In our experimental proposals, these paths can be either actually spatially separated paths, as in Fig. \ref{fig:interferometer}, or can represent two different internal states of the particle. Let us first consider the case of an electron propagating in a Mach-Zehnder interferometer, as depicted in Fig. \ref{fig:interferometer}. Later we will discuss a proposal where the paths are the internal states of atoms from a Bose-Einstein condensate. During our discussion, only the $z$ component of the particles wave function will be considered, since the plane of propagation is the $x-y$ plane (see Fig. \ref{fig:interferometer}), while the force acts only in the $z$ direction.

First, the electron propagation path is split by a beam splitter BS$_1$ into two paths A and B. Path A is free from the influence of any force, while on path B an external electric field generates a force in the $\bs{\hat{z}}$ direction. The paths are re-joined at a second beam splitter BS$_2$, where they are mixed towards exit ports C and D. We will consider that the first beam splitter has a reflection coefficient $ir$ (with $r$ real) and a transmission coefficient $t=\sqrt{1-r^2}$, while the second one has reflection and transmission coefficients equal to $i/\sqrt{2}$ and $1/\sqrt{2}$ respectively. If $|\Phi\rangle$ is the prepared initial momentum state of the electron, then after the first beam splitter the state becomes
\begin{align}
|\Phi_1\rangle=t|\Phi,A\rangle+ir|\Phi,B\rangle,\label{bs1}
\end{align}
where $|\Phi,j\rangle$ stands for the electron propagating through path $j$ with the state $|\Phi\rangle$ representing the quantum state of the $z$ component of the electron wave function.

After BS$_1$ each component A and B will evolve differently, until they reach the second beam splitter BS$_2$. Due to the different propagation lengths and interactions associated with each path, the electron state inside the interferometer just before BS$_2$ is:
\begin{align}
|\Phi_2\rangle=t|\Phi,A\rangle+ire^{i\beta}|\Phi',B\rangle. \label{cap}
\end{align}
where $|\Phi',B\rangle$ stands for the change in the momentum state of the component traveling through path B, and $\beta$ is the phase difference between the arms due to propagation.

Finally, after BS$_2$ we will have on exit ports C and D the respective (non-normalized) states
\begin{align}
|\Phi_C\rangle=\frac{t}{\sqrt{2}}|\Phi\rangle-\frac{re^{i\beta}}{\sqrt{2}}|\Phi'\rangle,\label{c}\\
|\Phi_D\rangle=\frac{t}{\sqrt{2}}|\Phi\rangle+\frac{re^{i\beta}}{\sqrt{2}}|\Phi'\rangle.\label{d}
\end{align}
We keep the $1/\sqrt{2}$ factors in the above expressions so that $\int|\langle p|\Phi_C\rangle|^2dp+\int|\langle p|\Phi_D\rangle|^2dp=1$, where $p$ is the component of the electron momentum in the $z$ direction. 

In our analysis, we consider that the external force $\mb{F}$ displaces the electron $z$ momentum wave function by a positive quantity $\delta$ without changing its form. This can be achieved with an impulsive force, characterized by a potential acting during a very short period of time \cite{feynman-path}. In this case, the evolved electron state after its propagation through path A is ${\langle p|\Phi'\rangle\approx e^{i\gamma}\Phi(p-\delta)}$, where $\Phi(p)$ is the initial electron wave function and $\gamma$ represents a possible extra phase.

Finally, we can write the wave functions corresponding to the states of Eqs. \eqref{c} and \eqref{d}, respectively, as
\begin{align}
\Phi_C(p)=\langle p|\Phi_C\rangle=\frac{t}{\sqrt{2}}\Phi(p)-\frac{re^{i\alpha}}{\sqrt{2}}\Phi(p-\delta),\label{cwf}\\
\Phi_D(p)=\langle p|\Phi_D\rangle=\frac{t}{\sqrt{2}}\Phi(p)+\frac{re^{i\alpha}}{\sqrt{2}}\Phi(p-\delta),\label{dwf}
\end{align}
with $\alpha = \beta + \gamma$ and $r=\sqrt{1-t^2}$.


We can now investigate how the average $z$ momentum of the final wave functions associated with exit ports C and D relate to that of the initial wave function $\Phi(p)$. We will specify $\Phi(p)$ as a Gaussian function with width $W$ and zero mean value,
\begin{align}
\Phi(p)=\frac{\pi^{-\frac{1}{4}}}{\sqrt{W}}\ \text{exp}\left[-\frac{1}{2}\left(\frac{p}{W}\right)^2\right].\label{gauss}
\end{align}
To maximize the anomalous force effect, we will fix ${\alpha=2n\pi}$ with integer $n$ in Eqs. \eqref{cwf} and \eqref{dwf}, which can be achieved by adjusting the relative $A$ and $B$ path lengths.

\begin{figure}[t]
    \centering \includegraphics[width=8.0cm]{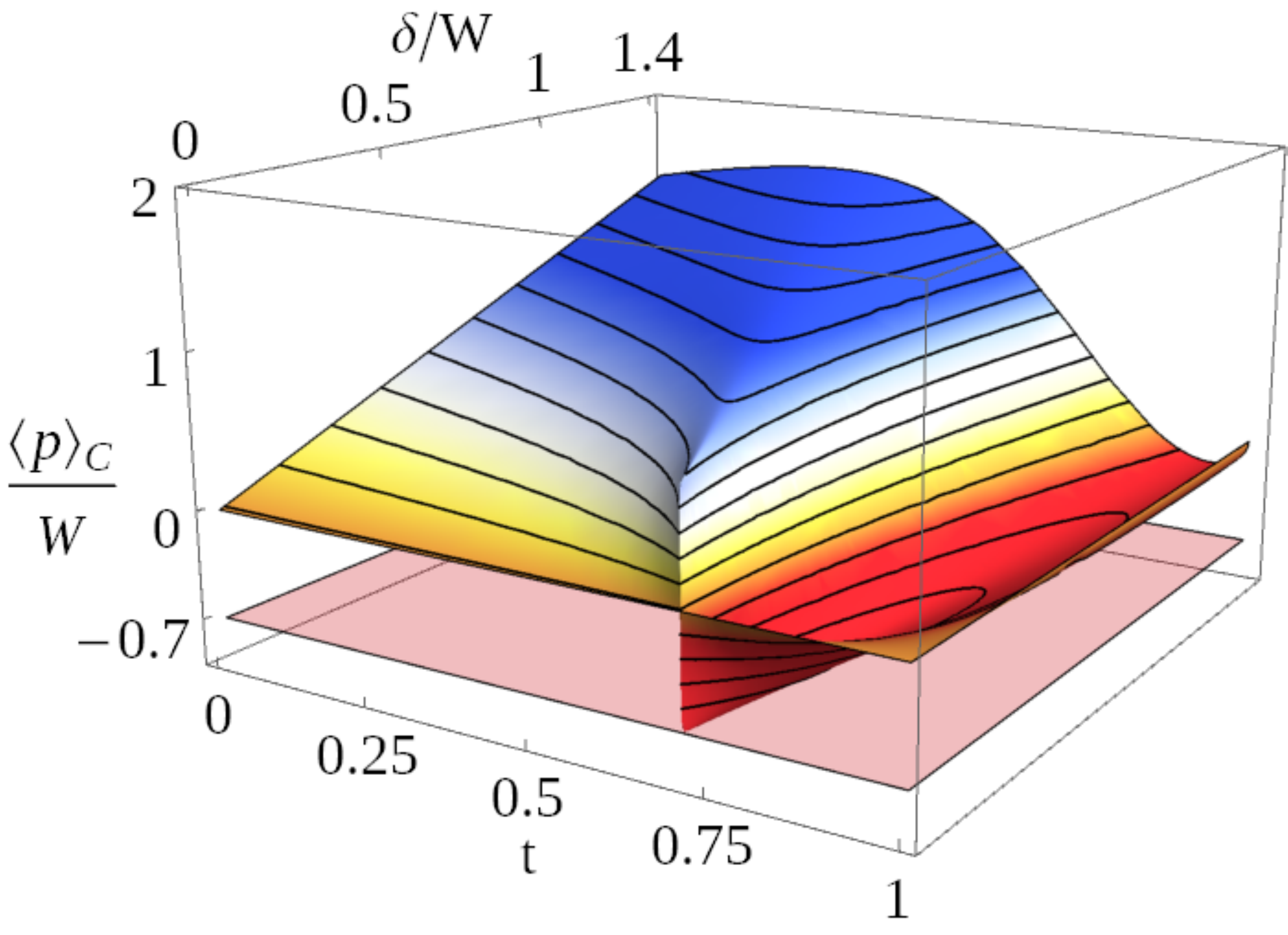}\\
  \caption{Average momentum $\langle p\rangle_C$ of the electrons selected at port C in units of the Gaussian width $W$, as a function of the BS$_1$ transmission coefficient $t$ and of the momentum kick $\delta$, also in units of $W$. The electron wave function is given by Eq. \eqref{cwf} with $e^{i\alpha}=1$ and $\Phi(p)$ from Eq. \eqref{gauss}. The red region is where $\langle p\rangle_C < 0$. The transparent plane corresponds to $\langle p\rangle_C=-0.7W$.}\label{fig:momentum}
\end{figure}

The average momentum of the wave function at a port $j$ is ${\langle p\rangle_j = (1/P_j)\int p|\Phi_j(p)|^2 dp}$, where ${P_j=\int |\Phi_j(p)|^2 dp}$ is the probability that the electron will exit through port $j$. In Fig. \ref{fig:momentum} we plot the average momentum $\langle p\rangle_C$ of the electrons selected at port C in units of the Gaussian width $W$, as a function of the BS$_1$ transmission coefficient $t$ and of the momentum kick $\delta$ in arm B, which is also in units of $W$. It is clear that there is a range of parameters for which the average momentum is negative, and it can attain values as negative as $-0.7W$. These values are clearly out of the weak interaction regime considered in Ref. \cite{aharonov132}, for we see that neither $\delta$ nor $|\langle p\rangle_C|$ have to be much smaller than $W$ in order to get $\langle p\rangle_C<0$.

So we see that, for an ensemble of quantum particles, the combination of a force in the positive $z$ direction with a null force may generate a displacement of the average momentum of these particles in the negative $z$ direction, when the appropriate post-selection is made. In other words, the superposition of a positive force with a zero force may generate a ``negative force'' on quantum particles. This is a counter-intuitive behavior with no classical analogue. For classical particles, any momentum-independent post-selection must generate either positive or zero average momentum. This odd feature can be explained as an interference effect.

\begin{figure}[t]
    \centering \includegraphics[width=6.5cm]{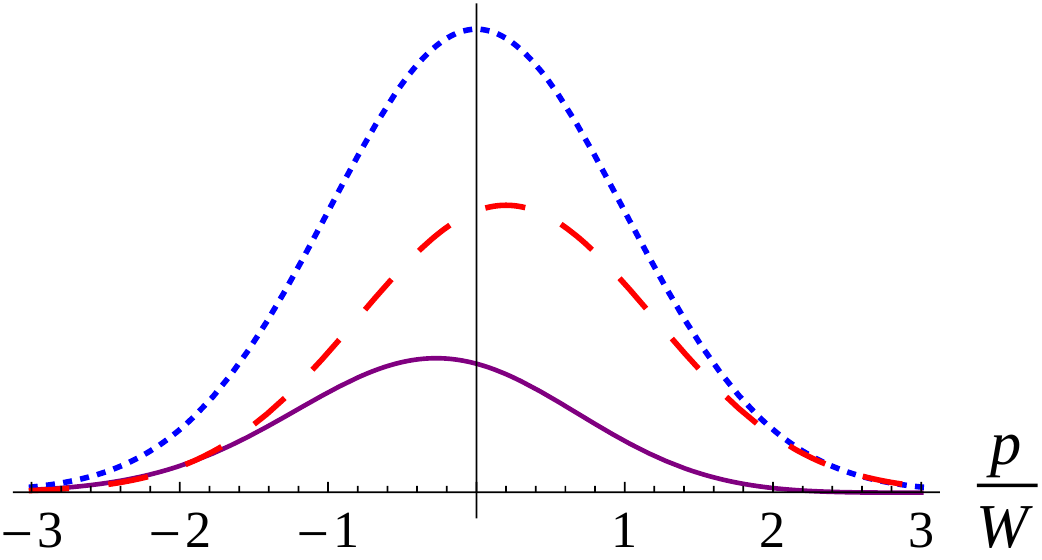}\\
  \caption{Components of the wave function of Eq. \eqref{cwf} when $\Phi(p)$ is given by the Gaussian function of Eq. \eqref{gauss}, with $t=0.85$, $\delta=0.2W$, and $e^{i\alpha}=1$: $t\Phi(p)/\sqrt{2}$ (blue dotted curve), $r\Phi(p-\delta)/\sqrt{2}$ (red traced curve), and $\Phi_C(p)$ (continuous purple curve).} \label{fig:specific}
\end{figure}

We plot in Fig \ref{fig:specific} the components $t\Phi(p)/\sqrt{2}$ and ${r\Phi(p-\delta)/\sqrt{2}}$ of Eq. \eqref{cwf} when $\Phi(p)$ is given by Eq. \eqref{gauss}, for $t=0.85$ and $\delta=0.2W$. The resulting wave function $\Phi_C(p)$ is the difference between these two components, as seen in Eq. \eqref{cwf} when $e^{i\alpha}=1$. In Fig. \ref{fig:specific}, $\Phi_C(p)$ is also plotted, and it has an average momentum around $-0.3W$. We can see that the positive momentum part of the displaced function $r\Phi(p-\delta)$ subtracts more from the positive than from the negative momentum part of $t\Phi(p)$, leaving an amplitude that is higher on the negative than on the positive momentum part of $\Phi_C(p)$. This is why the superposition of a positive-mean with a zero-mean function can result in a negative-mean one. It is a destructive interference effect on momentum wave functions, which is why we call it an \emph{interference of force}. It is also clear that no such anomalous force occurs if the displacement $\delta$ is larger than the wave function width $W$, since the interference would be negligible in this case. But it is not necessary that $\delta\ll W$, as considered in Ref. \cite{aharonov132}. When $\delta$ is smaller than $W$, but of the same order of magnitude, as in Fig. \ref{fig:specific}, the anomalous force is maximized while the final wave function keeps a considerable amplitude, which means this case should be easier to be observed in an experiment.  

In general, quantum particles suffer diffraction through the interaction with energy potentials, exchanging momentum with the agent of that potential. However, when the gradient of the potential does not vary much in the region occupied by the particle wave function, Ehrenfest's theorem tells us that the average momentum gained by the particle should be the expected classical one, computed as the result of a classical force acting on the particle \cite{cohen}. The component of the wave function that propagates through path B in the interferometer of Fig. \ref{fig:interferometer} does receive an average momentum equal to the classically computed one. But when this component interferes with the component that propagates through path A with no momentum exchange, the average momentum received by the particle when it exits by port C can be in the opposite direction in relation to the classically expected one, as depicted in Figs. \ref{fig:momentum} and \ref{fig:specific}. This interference of different momentum exchanges, which may result in a momentum displacement that corresponds to a force applied in the opposite direction, is a genuinely quantum effect with no classical analogue.


\begin{figure}[t]
    \centering \includegraphics[width=8.0cm]{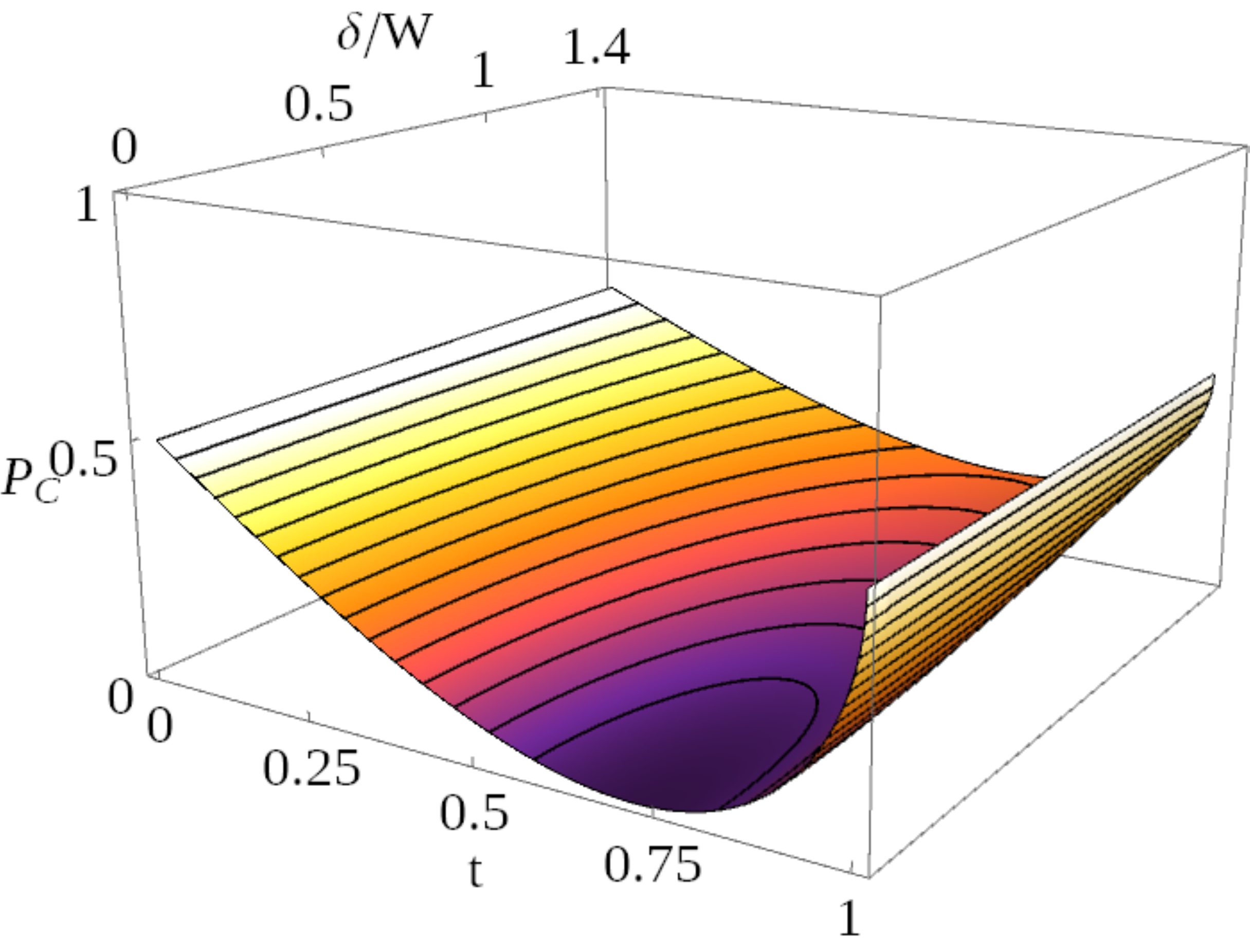}\\
  \caption{Probability $P_C$ that an electron exits the interferometer of Fig. \ref{fig:interferometer} at port C, as a function of the transmission coefficient $t$ of BS$_1$ and of the kick $\delta$ in units of the Gaussian width $W$. The electron wave function is given by Eq. \eqref{cwf} with $e^{i\alpha}=1$ and $\Phi(p)$ from Eq. \eqref{gauss}.}\label{fig:probability}
\end{figure}

\begin{figure}[t]
    \centering \includegraphics[width=8.0cm]{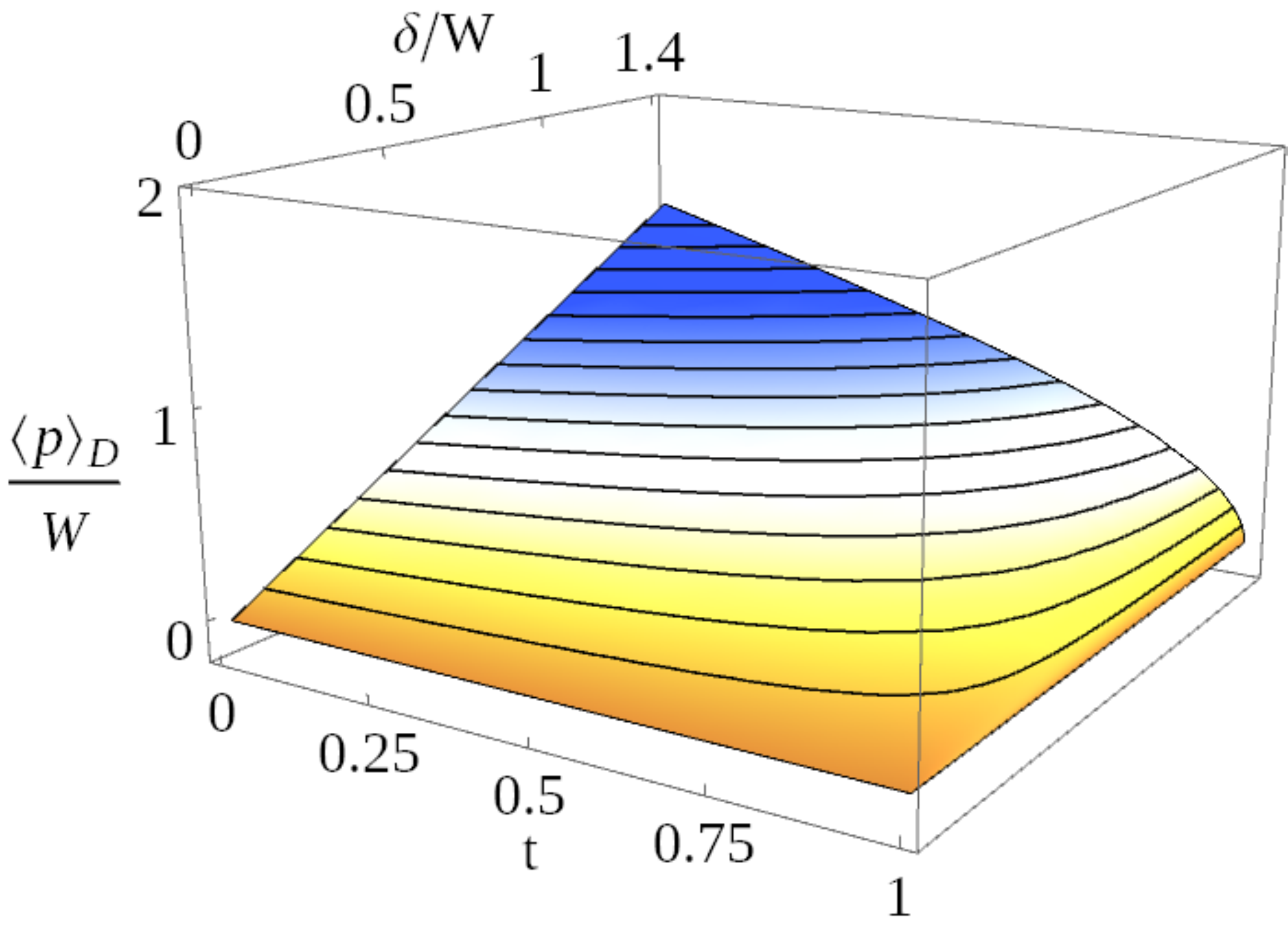}\\
  \caption{Average momentum $\langle p\rangle_D$ of the electrons selected at port D in units of the Gaussian width $W$, as a function of the BS$_1$ transmission coefficient $t$ and of the momentum kick $\delta$, also in units of $W$. The electron wave function is given by Eq. \eqref{dwf} with $e^{i\alpha}=1$ and $\Phi(p)$ from Eq. \eqref{gauss}.}\label{fig:momentumD}
\end{figure}

One could wonder if the described process suggests that momentum is not being conserved. Or, stated in another way, how could we get the electron with average momentum in a direction if it has only gained momentum in the opposite one? In order to understand this we have to stress that, because the negative momentum transfer takes place with destructive interference, the successful selection of the electron at the interferometer port C must be unlikely. For instance, in the case illustrated in Fig. \ref{fig:specific} the probability $P_C$ that the electron comes out at port C is around $6\%$. To have a quantitative view of that, we plot in Fig. \ref{fig:probability} $P_C$ as a function of $t$ and $\delta/W$. By comparison with Fig. \ref{fig:momentum}, clearly the region with most negative average momenta coincides with the lowest probabilities of successful selection. Let us take a look at the discarded events. If we compute the momentum average $\langle p\rangle_D$ of the case that the electron is selected on port D of the interferometer, we see that the whole range of average momenta is positive (Fig. \ref{fig:momentumD}). 
When we sum both averages weighed by their respective probabilities, we see that
\begin{align}
P_C\langle p\rangle_C + P_D\langle p\rangle_D = t^2\langle p\rangle_{\Phi} + r^2\langle p\rangle_{\Phi'},
\end{align}
where $\langle p\rangle_{\Phi}$ and $\langle p\rangle_{\Phi'}$ are the momentum averages calculated with the corresponding states $|\Phi,A\rangle$ and $|\Phi',B\rangle$ of Eq. \eqref{cap}. This is exactly what we would expect for conservation of momentum, since $r^2$ and $t^2$ are the respective probabilities that the electron has or has not interacted with the external field, such that the Ehrenfest theorem is not violated. What the post-selection allows us to do, then, is a re-arrangement of the probabilities of selecting the electron with certain transverse momentum eigenvalues. We correlate most of the amplitude for positive $z$ momentum with the electron exiting through port $D$, and leave the negative values associated with the electron exiting at $C$. The impression that momentum is not conserved then comes from the fact that we are discarding many detection events. 

Matter interferometry is an extensively developed field, with decades of research on electron, neutron and atom interferometers \cite{marton54,rauch74,keith91,cronin09}. Mach-Zehnder interferometers with electrons in free space can be constructed with diffraction gratings acting as beam splitters and mirrors \cite{godun01,gronniger06}. It should not be difficult to place a capacitor in one of the arms of the interferometer, since this was already done with similar atomic interferometers \cite{ekstrom95,roberts04,miffre06}. A thin metallic foil should be placed between the interferometer arms near the capacitor to avoid the presence of the capacitor field on path A of Fig. \ref{fig:interferometer}. The gratings with ${100\ \T{nm}}$ periodicity used in Ref. \cite{gronniger06}, when illuminated by their ${6\ \T{keV}}$ electron beam source, can generate a separation of ${55\ \mu\T{m}}$ between the two paths at a distance of ${35\ \T{cm}}$ from the grating. In Ref. \cite{ekstrom95}, which performs the interferometer with atoms, this separation is enough to have control of different capacitor tensions for each beam path. By using the slit of ${1.5\ \mu\T{m}}$ of Ref. \cite{gronniger06} for the initial electron spatial state preparation, we approximate it by a Gaussian profile and estimate the beam width after ${1\ \T{m}}$ of propagation to be ${1.7\ \mu\T{m}}$, such that the wave function spreading is small. In this setup, for capacitor plates separated by ${1\ \T{mm}}$ with a length of ${1\ \T{cm}}$ in the beam propagation direction and with an applied tension of ${0.2\ \T{mV}}$, the displacement of the momentum wave function is around 10\% of its width, compatible to the displacement needed to show the anomalous force effect. So the experimental verification of the effect should be feasible with current technology.

It should also be possible to perform the experimental verification of the quantum interference of force with neutron interferometers \cite{rauch-neutron,hasegawa11}. Nowadays it is possible to apply different magnetic fields on each arm of a Mach-Zehnder neutron interferometer \cite{geppert14,denkmayr17}. If this setup is adapted to apply an inhomogeneous magnetic field in one of the interferometer arms, producing a Stern-Gerlach force, and no field on the other, it should be possible to produce the superposition of a positive force with a null force on the neutrons resulting in an average negative momentum transfer to them with the appropriate post-selection.  

Another possibility for the experimental verification of the quantum interference of force is in a setup with atoms from a Bose-Einstein condensate when the atomic trap is turned off. In this case, paths A and B of Fig. \ref{fig:interferometer} can be associated to internal atomic states $|A\rangle$ and $|B\rangle$ with an energy difference on the microwave region and definite magnetic moments in the $z$ direction. If the atoms are all initially in the state $|\Phi(p)\rangle|A\rangle$,  where $|\Phi(p)\rangle$ represents the quantum state of the $z$ component of the atoms wave function in momentum space, a microwave pulse resonant with the transition can create the superposition $t|\Phi(p)\rangle|A\rangle+\sqrt{1-t^2}|\Phi(p)\rangle|B\rangle$, where $t$ (assumed real) depends on the duration and amplitude of the microwave pulse. The subsequent application of an inhomogeneous magnetic field that generates a Stern-Gerlach force during a short period of time can create the state $t|\Phi(p-\delta_a)\rangle|A\rangle+\sqrt{1-t^2}{|\Phi(p-\delta_b)\rangle}|B\rangle$, where $\delta_a$ and $\delta_b$ depend on the duration and spatial configuration of the inhomogeneous magnetic field and on the $z$ component of the magnetic moments of the states $|A\rangle$ and $|B\rangle$ respectively. A microwave resonant $\pi/2$ pulse can then transform this state into ${t|\Phi(p-\delta_a)\rangle}[|A\rangle+|B\rangle]/\sqrt{2}+\sqrt{1-t^2}{|\Phi(p-\delta_b)\rangle}[-|A\rangle+|B\rangle]/\sqrt{2}$. The application of a second inhomogeneous magnetic field, but now transmitting the opposite momentum in relation to the previous one, followed by a selection of the atoms in the atomic  state $|A\rangle$, generates the momentum state $t/\sqrt{2}|\Phi(p)\rangle -  \sqrt{1-t^2}/\sqrt{2}|\Phi(p-\delta_b+\delta_a)\rangle$. This state is associated to the same momentum wave function as Eq. \eqref{cwf} with $\rme^{i\alpha}=1$ and $\delta=\delta_b-\delta_a$ and can thus show the anomalous force effect. The combination of a total positive momentum transfer with a total zero momentum transfer to the atoms may result in a negative average momentum transfer.

In summary,  we have shown how the quantum superposition of a positive force on a quantum particle with no force on the particle may result in a ``negative force'' on that particle, in a phenomenon that we named \textit{quantum interference of force}. We also presented proposals for feasible experiments that could verify this effect with current technology. The quantum interference of force can generate rather curious effects, such as an effective attraction between electrons in an interferometer when we post-select by which exit each electron leaves the interferometer \cite{cenni18}.


The authors acknowledge S\'ergio Muniz and Gustavo Telles for very useful discussions. This work was supported by the Brazilian agencies CNPq, CAPES and FAPEMIG.



%

\end{document}